\documentclass[showpacs, amsmath,amssymb]{revtex4}
\usepackage{amssymb}
\usepackage[dvips]{graphicx}

\begin{document}

\title{Spin excitations in systems with hopping electron transport and strong position disorder in a large magnetic field.}
\author{A. V. Shumilin}

\affiliation{A.F.Ioffe Physico-Technical Institute, St.-Petersburg 194021, Russia.}

\begin{abstract}
We discuss the spin excitations in systems with hopping electron conduction and strong position disorder. We focus on the problem
in a strong magnetic field when the spin Hamiltonian can be reduced to the effective single-particle Hamiltonian and treated with
conventional numerical technics. It is shown that in a 3D system with Heisenberg exchange interaction the spin excitations have a delocalized part of the spectrum even in the limit of strong disorder, thus leading to the possibility of the coherent spin transport. The spin transport provided by the delocalized excitations can be described by a diffusion coefficient. Non-homogenous magnetic fields lead to the Anderson localization of spin excitations while anisotropy of the exchange interaction results in the Lifshitz localization of excitations. We discuss the possible effect of the additional exchange-driven spin diffusion on the organic spin-valve devices.
\end{abstract}


\maketitle

\section{Introduction}

The spin transport in systems with hopping conduction attracts a significant attention in the last years. It is in particular related to the recent progress in the organic spintronic devices \cite{dediu2002,vardeny2004, vert1, vert2}.  A strong magnetoresistance was found in ${\rm LSMO/T_6/LSMO}$ planar structures in \cite{dediu2002}.  Vertical organic devices with spin-valve magnetoresistance were studied in \cite{vardeny2004, vert1, vert2}. In these devices spin-polarized electrons are injected from a ferromagnetic contact into the normal organic material. The transport in organics is supposed to be related to the hopping of polarons \cite{Basler_polarons}. The injected spin is detected by the second ferromagnetic contact. The conductivity of the device depends on the relative orientation of the magnetizations of the ferromagnetic contacts. The vertical organic spin-valves show a number of non-trivial phenomena like the negative sign of the spin-valve effect \cite{negMR} and the memristor effect \cite{memristor}.

However in spite of extensive experimental and theoretical studies  the physics of the organic spin-valves is far from being well-understood. The most puzzling problem is perhaps the failure to observe the Hanle effect in organic spin-valves. The small external magnetic field applied perpendicularly to the axis of the contact magnetization should lead to a spin precession inside the organic layer and dramatically affect the magnetoresistance. However, no effect of the perpendicular magnetic field on the organic spin-valves was reported \cite{no-Hanle-1,schmidt_hanle}.

One of the most important parts in the description of the organic spin-valves is the spin transport inside the organic layer.
It is often assumed that the spin transport in materials with hopping conduction occurs due to hops of spin-polarized electrons and is related to the charge transport. This mechanism of spin transport is relatively well-understood \cite{Bobbert-spin-dif,Kin} and can be described for example by kinetic equations. However, a novel idea was proposed in the recent study of Z.G. Yu \cite{Yu}. It was assumed that the spin transport can be decoupled from the charge transport and occur due to the exchange interaction of the localized electrons. It is stated in \cite{Yu} that sometimes the spin diffusion can be many orders of magnitude faster than the charge diffusion. The absence of the Hanle effect in the organic spin-valves was then attributed to this fast spin diffusion \cite{Yu2}. The hypothesis of this decoupling is supported by recent observations of the pure spin transport in ${\rm Alq_3}$ \cite{pure-spin}.

Another reason to be interested in the spin transport in systems with hopping conduction is the recent progress in studies of the spin noise in systems with localized electrons such as doped semiconductors and quantum dot arrays. Both experimental \cite{noise-QD-1,noise-QD-2} and theoretical \cite{Gl-Iv-PRB, Gl-Iv-FTT, Glazov-spin-noise} results  appeared recently in this field. It was shown that the noise spectrum is sometimes closely related to the spin motion \cite{Glazov-spin-noise}. Therefore, understanding of the spin transport in semiconductors with hopping conductivity can be important for the analysis of the spin noise in these materials.

We would like to note that the main assumptions in \cite{Yu} are related to the exponential dependence of the exchange integrals on the distance with the localization length $a \sim 1\,nm$ and random positions of electrons with concentrations $n \sim 10^{17} - 10^{19}\, cm^{-3}$. These assumptions are applicable both to organic materials and to doped semiconductors. No suggestions explicitly specific to organic materials were made in \cite{Yu}.

The theory \cite{Yu} is based on the Redfield equation \cite{Redfield}. This equation describes the weak coupling of the system to some external degrees of freedom. In the discussed problem the relevant degrees of freedom are  phonons and the Redfield equation is adequate for the description of phonon-induced electron hops. However, the spin-phonon coupling is very weak in these systems. The spin-phonon interaction times for localized states in $GaAs$-based structures were discussed in \cite{Frenkel-S-Ph,S-Ph-2} and were found to be as large as $10^{-5} - 10^{-1}\, s$ at least at low temperatures. The dominant mechanism of spin-phonon interaction is related to the spin-orbit interaction, and is suppressed in the organic materials \cite{SO_inO, Yu-SO2}. The times of spin-lattice relaxation in organic measured in spin resonance experiments are somewhat smaller $\sim 1 \mu s$ \cite{EPR-rel1,EPR-rel2}. However, it is not clear if these times reflect the direct spin-phonon coupling or the spin-relaxation due to the hopping and the spin-orbit interaction. In the present study, we believe that spin-phonon interaction cannot significantly affect the spin transport. However, it is possible that spin transport due to the exchange interaction occurs without interaction with phonons in a coherent way. It is important that the possibility of the phonon-independent spin transport due to the exchange interaction cannot be studied with a perturbation-based approach (like the Redfield equation). Naturally, the perturbation theory for the conventional hopping is based on the assumption that the overlap integrals are small compared to the large random electron energy or the polaron energy. These large energies do not affect the phonon-independent spin transport related to the exchange interaction. Therefore all the relevant energies in the system are related to the exchange interaction. We believe that in this case the spin transport should be studied not with the Redfield equation but by the analysis of the excitations in the spin Hamiltonian.

The coherent spin transport is possible when spin excitations are delocalized (even if the charge excitations are localized). Let us note that the localization of the charge excitations is closely related to the fact that random electron energies ${\cal E}_{el}$ are larger than overlap integrals $t_{el}$ leading to the Anderson localization of charges. The spin-flip processes do not change the positions of charges and are not related to ${\cal E}_{el}$ and to the conventional parameter of the Anderson localization. However, the distribution of the exchange integrals in the discussed systems is exponentially-broad. It makes the problem of coherent spin transport in hopping systems in some sense similar to the problem of Lifshitz localization \cite{Lifshitz, Efr-Shk}, although this similarity is not rigorous.
 Therefore to understand the possibility of  the coherent spin transport one should directly study the spin excitations in the media with an exponentially-broad distribution of the exchange integrals.

We are not aware of previous studies considering the problem of spin excitations in a system with exponentially-broad distribution of the exchange integrals. In the present work, we consider this problem in one specific case when a high external magnetic field is applied to the system.  In this case, the spin excitation problem can be reduced to effective one-particle Hamiltonian. This effective Hamiltonian can be treated with conventional numeric technics developed to study the electron localization. In the present study, we apply the method of twisted boundary conditions \cite{ThouOr,Thou2}. We show that for pure-Heisenberg exchange interaction some excitations are localized but in 3D there is always a band of delocalized excitations corresponding to some diffusion coefficient. These excitations can provide a coherent phonon-independent transport mechanism.
Random magnetic or hyperfine fields lead to the Anderson localization and the anisotropic exchange interaction lead to a localization similar to one in the Lifshitz problem \cite{Lifshitz, Efr-Shk}. In both  cases, the coherent spin transport over macroscopic distances is impossible for sufficiently small $na^3$.

The paper is organized as follows. In section \ref{sect2} we discuss the Hamiltonian of a disordered spin system with Heisenberg interaction and reduce it to the effective one-particle Hamiltonian in the case of a strong magnetic field. In section \ref{sect3} we discuss the numerical methods used in our study. In section \ref{sect4} we apply these methods to the Heisenberg exchange interaction problem. We show that some of the spin excitations are delocalized in this problem regardless of the disorder. These excitations can be described by a diffusion coefficient that is larger than the charge diffusion coefficient for realistic parameters. In section \ref{sect5} we modify the Hamiltonian to include the effective random magnetic fields and the spin-orbit interaction. We show that these effects can lead to the localization of all the spin excitations. In section \ref{sect6} we give a qualitative explanation for the localization and delocalization picture obtained in the numeric simulations. In section \ref{sect7} we discuss the effect of the delocalized spin excitations on the spin transport. In section \ref{sect8} we give a short conclusion of our results.

\section{Reduction to the one-partical problem}
\label{sect2}

We consider a system of localized electrons with Heisenberg exchange interaction in an external magnetic field. The Hamiltonian of the system is
\begin{equation} \label{H_0}
H = g \mu_B {\bf B} \sum_i {\bf S}_i + \sum_{ij} J_{ij} {\bf S}_i {\bf S}_j.
\end{equation}
Here {\bf B} is the external magnetic field, $\mu_B$ is Bohr magneton, and $g$ is $g$-factor. ${\bf S}_i$ is the operator of the spin of $i$-th electron. We neglect electron motion and each spin ${\bf S}_i$ has a well-defined position ${\bf r_i}$. We consider these positions to be uncorrelated and randomly distributed over 3D space.

The exchange interaction is assumed to have Heisenberg form (the anisotropic exchange interaction is discussed later). The exchange energies  $J_{ij}$ exponentially decay with the distance between electrons $i$ and $j$, $r_{ij} = |{\bf r}_i - {\bf r}_j|$ \cite{Exc-Int}
\begin{equation}\label{J_ij}
J_{ij} = J_0  \left( \frac{r_{ij}}{a} \right)^{5/2} \exp(-2r_{ij}/a).
\end{equation}
Here $J_0=0.82e^2/\epsilon a$, $a$ is the electron localization length and $\epsilon$ is the dielectric susceptibility. We consider the case of small electron concentration $na^3 \ll 1$ and the distribution of  the exchange energies (even between neighbors) is exponentially-broad.

The Hamiltonian (\ref{H_0}) conserves the total spin projection on  the axis $z$ of the external magnetic field ${\bf B}$. Let us assume that the initial state of the system had a definite spin projection $S_z^{(0)}$. Then one of the spins is flipped so that z-projection of the total spin becomes $S_z^{(0)} + 1$. The dynamics of the additional magnetic moment is closely related to the eigenstates of the Hamiltonian (\ref{H_0}) corresponding to the z-projection of the total spin $S_z^{(0)} + 1$. The number of these eigenstates in the general case is very large. For the system with $N$ sites with total spin $S_z$ the number of states is
\begin{equation}
N_{states} = \frac{N!}{K!(N-K)!}, \quad K = \frac{N}{2} - S_z,
\end{equation}
where $K$ is the number of electrons with spin up. It makes the numerical analysis to be rather complicated.

However, there is a special case when it becomes much more simple. Let us assume that the external magnetic field $B$ is rather strong and the initial state corresponds to the case when all electron spins are aligned along the magnetic field $S_z^{(0)} = -N/2$. After the excitation, the total spin projection is $-N/2+1$. As we mentioned before it does not relax due to the action of Hamiltonian (\ref{H_0}). The spin relaxation to the initial state is possible only due to the weak spin-phonon interaction. We consider the system at times that are less than the time of this relaxation. Therefore, we are interested only in the eigenstates with the total spin projection $-N/2+1$. The number of these states is $N$. It means that we can apply conventional numerical technics for sufficiently large numerical samples.

Let us show that in the considered case our problem can be reduced to an effective one-particle problem with some effective one-partical Hamiltonian. Let us apply the Jordan-Wigner transformation \cite{Levitov-Shitov}.
\begin{equation}\label{Jor-Wig}
\sigma_i^z = 2a_i^+a_i-1, \quad \sigma_i^+ = a_i^+ \prod_{k<i}\sigma_k^{z}, \quad \sigma_i^- = a_i \prod_{k<i} \sigma_k^z,
\end{equation}
where $\sigma$ are the Pauli matrixes related to spin operators as usual ${\bf S} = \frac{1}{2}{\boldsymbol \sigma}$,  $\sigma_{i}^\pm = \sigma_i^x \pm \sigma_i^y$. The index $i$ numerates the electron spins in an arbitrary order. $a_i^+$ and $a_i$ are the creation and the destruction operators of the effective particle corresponding to the spin $\sigma_i$. The transformation yields the fermion-like Hamiltonian
\begin{equation}\label{FermHam}
H = -\frac{1}{2}\sum_{i>j} J_{ij} (a_i^+ a_j + a_j^+ a_i) \prod_{i>k>j} \sigma_k^z +
\end{equation}
$$
+ \frac{1}{4}\sum_{i>j} J_{ij} (2n_i - 1) (2 n_j -1) + \sum_i \frac{g \mu_b B}{2} (2n_i-1).
$$
In the general case it differs from the Hamiltonian of a fermion problem due to the product $\prod_{i>k>j} \sigma_k^z$. However when the total spin projection is equal to $-N/2+1$ these products are reduces to constants.

Let us consider the Hamiltonian (\ref{FermHam}) in the basis corresponding to this spin projection. We chose the basis as follows. Wavefunction $\psi_i$ corresponds to the situation when the flipped spin is positioned on electron $i$ while the spins of other electrons are directed along the magnetic field. In this case when we consider the matrix element $H_{ij} = \langle \psi_i |H|\psi_j \rangle$ we should take spins on all the sites $k\ne i,j$ to be directed along the magnetic field. It allows us to deal with the products $\prod_{i>k>j} s_k^z$ and reduce the problem to a one-particle one. In the discussed basis the Hamiltonian has the form
\begin{equation}\label{Hmatrix}
H_{ij} = J_{ij}/2, \quad H_{ii} = E_0 - \sum_{j\ne i} J_{ij}/2
\end{equation}
where $E_0 = - g \mu_b B (N-1)/2 + (1/4) \sum_{i>j} J_{ij}$. $E_0 - g \mu_b B/2$ is the energy of the ground state of the system when all spins are polarized.

Let us note that the Hamiltonian (\ref{Hmatrix}) is quite similar to the Hamiltonian of the Lifshitz localization problem \cite{Lifshitz, Efr-Shk}
\begin{equation} \label{HLif}
H_{ij} = t_{ij}, \quad H_{ii} = 0.
\end{equation}
This problem deals with an electron on a set of sites (that can correspond for example to the impurity states) without random energies but with random positions and with overlap integrals $t_{ij} = t_0 \exp(-r_{ij}/a)$ that exponentially decrease with the distance $r_{ij}$ between the sites. This dependence leads to the exponentially-broad distribution of the non-diagonal Hamiltonian elements $H_{ij}$ that is similar to the distribution of non-diagonal elements in the Hamiltonian (\ref{Hmatrix}). The main difference between Hamiltonians (\ref{Hmatrix}) and (\ref{HLif}) are the diagonal elements. In the Lifshitz localization problem, the diagonal Hamiltonian elements are zero while in the spin problem they are equal to $- \sum_{j\ne i} J_{ij}/2$. These elements appear due to $\sigma_i^z \sigma_j^z$ term in the exchange interaction. In terms of effective particle, these elements correspond to on-site energy that is related to the exchange interaction with non-flipped spins.

The results of \cite{Lifshitz, Efr-Shk} are as follows. Even without energy disorder all the electron states of the Hamiltonian (\ref{HLif}) are localized when the parameter $na^3$ is small enough. In what follows, we compare our problem with the Lifshitz localization and discuss its similarities and differences with the problem (\ref{Hmatrix}).

\section{Twisted boundary condition method}
\label{sect3}

The  method of twisted boundary conditions \cite{ThouOr, Thou2} was first proposed by Edwards and Thouless and is a well-established method to study the electron localization. As long as our Hamiltonian is  equivalent to some effective electron problem we can use it as well. The essence of the method is as follows. First, a numerical sample with a finite size $L$ and periodic boundary conditions is considered. The Hamiltonian is solved and the set of eigenenergies $E_n$ is found. Then the twisted boundary conditions are introduced. It means that each non-diagonal Hamiltonian element $H_{ij}$ is ascribed with a small phase shift
$H_{ij} \rightarrow H_{ij} e^{i\varphi}$
if the transition from site $i$ to site $j$ crosses the right border of the numerical sample. The backward transition crosses the left border of the sample and is ascribed with the phase $-\varphi$ so that the Hamiltonian stays hermitian.  The dependence of state energies $E_n$ on the phase $\varphi$ is quadratic for small $\varphi$.
\begin{equation} \label{Thou-Sn}
E_n(\varphi) = E_{n}(0) + G_n \varphi^2
\end{equation}
The coefficients $G_n$ are random values (with random sign) and the mean value $ \langle G_n^2 \rangle^{1/2} = g_T(E)$ in some part of the spectrum $(E-\Delta E, E+\Delta E)$ is considered.

It can be shown that the dependence of the so-called "level curvature`` $g_T(E)$ on the system size $L$ is closely related to the localization of states with energy $E$. Naturally, when the states are localized with some localization distance $l_{loc} < L$ the electron states can ``feel" the phase twist only when they are near the border of the numerical sample. And even in this case, the state ``feels" only one twist of the boundary condition on the left or on the right side. However without the second twist, it corresponds to a simple change of the basis. Thus a finite coefficients $G_n$ appear only due to the finite probability of a localized electron to cross the numerical sample and $g_T$ is proportional to $g_T(E) \propto \exp(-L/l_{loc})$.

In the opposite case of delocalized electrons the dependence $g_T(L)$ follows a power law. It is known \cite{Thou2} that in the case  of a diffusive transport the diffusion coefficient $D$ is related to the level curvature $g_T$ at large $L$ as
\begin{equation}\label{diffusion}
D = C\cdot \frac{1}{\hbar}\lim_{L\rightarrow \infty} g_{T}L^2,
\end{equation}
where $C$ is the coefficient of the order of unity.

In the present study, we consider 3D numerical samples containing from 10 to 1000 sites. Unless otherwise stated the distance is measured in units $n^{-1/3}$ so the sample with the size $L=10$ contains $L^3=1000$ sites. The energy spectrum is separated into 20 intervals containing the equal number of energy states. The squares of the coefficients $G_n$ are then averaged over these intervals to calculate $g_T(E)$. The dependence of level curvature $g_T$ over the sample size $L$ is studied. All the numerical results are averaged over at least $1000$ disorder configurations.
We take $C=1$ in the equation (\ref{diffusion}). It means that our results for the diffusion coefficient should be considered as order-of-magnitude estimates.

In appendix 1, we apply the twisted boundary conditions method to the problem of Lifshitz localization that is closely related to our problem and can be considered as a test for our numerical methods. It can be seen that at relatively high concentration of sites in the Lifshitz problem some electron states are localized and some are delocalized. However at smaller concentrations the level curvatures $g_T$ exponentially decrease with system site $L$ for all electron energies. It means that all electron states are localized when the parameter $n^{1/3}a$ is below the threshold value of the Lifshitz localization.

\section{Heisenberg exchange interaction}
\label{sect4}

Let us now apply the twisted boundary conditions method to the spin excitations described with the Hamiltonian (\ref{Hmatrix}). We start the discussion from the case when $n^{1/3}a=0.2$. Note that the  exponential part of the dependence of the non-diagonal elements of Hamiltonian (\ref{Hmatrix}) on the distance between sites is $\exp(-2r_{ij}/a)$ in contrast to the dependence $\exp(-r_{ij}/a)$ in the Hamiltonian of the Lifshitz problem (\ref{HLif}). Thus $a=0.2n^{-1/3}$ in the spin problem roughly corresponds to $a=0.1n^{-1/3}$ in the Lifshitz problem that is below the localization threshold (see appendix 1).

\begin{figure}[htbp]
    \centering
        \includegraphics[width=1.0\textwidth]{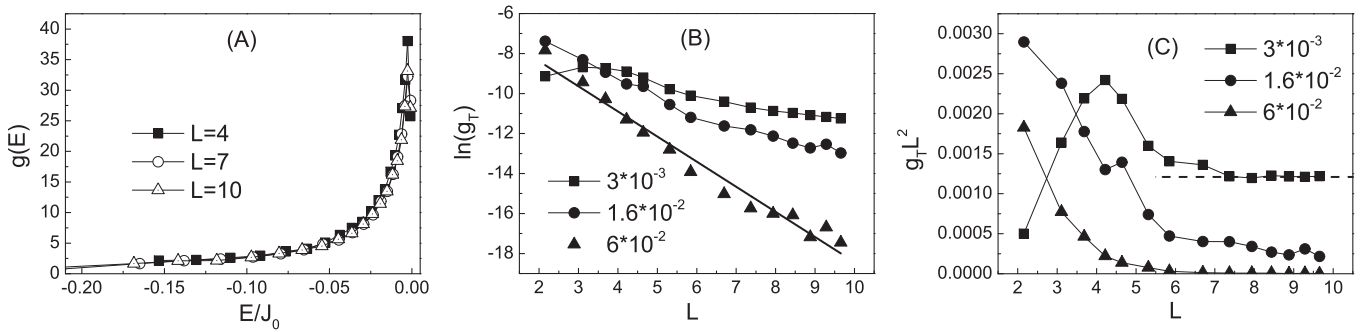}
        \caption{The density of states (a) and the dependence $g_T(L)$ (b, c) for the Hamiltonian (\ref{Hmatrix}) with $n^{1/3}a=0.2$. The numbers on plots (b) and (c) correspond  to the absolute values of the spin excitation energy $|E|/J_0$.    }
    \label{fig:Fer02}
\end{figure}

Figure \ref{fig:Fer02} shows the numerical results for the spin excitation problem with $n^{1/3}a=0.2$. The density of states is shown on fig. \ref{fig:Fer02}(A) for several numerical sample sizes $L$. Its dependence on $L$ quickly saturates at $L \gtrsim 4$. The zero energy on this plot corresponds to $E_0$ from eq. (\ref{Hmatrix}). It appears that all the excitation energies are less than $E_0$ therefore the density of states is zero at $E>0$. It has a sharp maximum near $E=0$ and a tail at $|E|>0.03J_0$.

The dependence of $g_T$ on the numerical sample size $L$ for different energies is shown at fig. \ref{fig:Fer02} (B) and (C).
We compared this dependence with localization and diffusion hypothesis.
For the localized states $g_T$ exponentially decreases with $L$, $g_T \propto \exp(-L/a_{loc})$. On fig. \ref{fig:Fer02} (B) we show the dependence $g_T(L)$ in the logarithmical scale. For the localized states the dependence $g_T(L)$ on this scale corresponds to the straight line. The localization distance can be derived from the slope of this line. On the figure \ref{fig:Fer02} (B) we show this asymptotic with a solid line for $E = 0.6 \cdot 10^{-2} J_0$. The localization distance derived from the slope of this line is $a_{loc} =0.8 n^{-1/3}$.  On fig. \ref{fig:Fer02} (C) we show the dependence of $g_T L^2$ on $L$. For the states corresponding to the diffusion $g_TL^2$ should tend to a non-zero constant at $L \rightarrow \infty$. This constant corresponds to the dimensionless diffusion coefficient. The estimated diffusion coefficient $D= 1.2 \cdot 10^{-3} J_0 n^{-2/3}/\hbar$ is shown on fig. \ref{fig:Fer02} (C) with a dashed line for the energy $E=3 \cdot 10^{-3} J_0$.

\begin{figure}[htbp]
    \centering
        \includegraphics[width=0.75\textwidth]{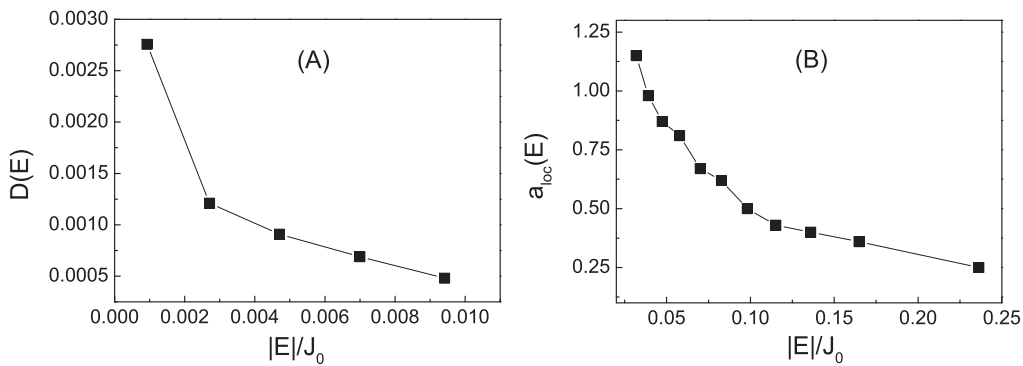}
        \caption{The diffusion coefficient (A) for delocalized states measured in units $J_0 n^{-2/3}/\hbar$ and the localization distance (B) in units $n^{-1/3}$ for localized states.  }
    \label{fig:DaLoc}
\end{figure}

The discussed situation is similar to the Lifshitz problem above the localization threshold: a part of the spectrum corresponds to localized excitations while the excitations near the maximum of the density of states are delocalized and can be described by a diffusion coefficient. The localization distances and the diffusion coefficients are different for different energies. On figure \ref{fig:DaLoc} we show the energy dependence of diffusion coefficient (for the delocalized states with $|E|/J_0 <0.01$) and for the localization length (for the localized states with $|E|/J_0 >0.03$). The states in the energy interval $0.01 <|E/J_0|<0.03$ cannot be reliably identified as localized or delocalized with our numerical data.

\begin{figure}[htbp]
    \centering
        \includegraphics[width=0.75\textwidth]{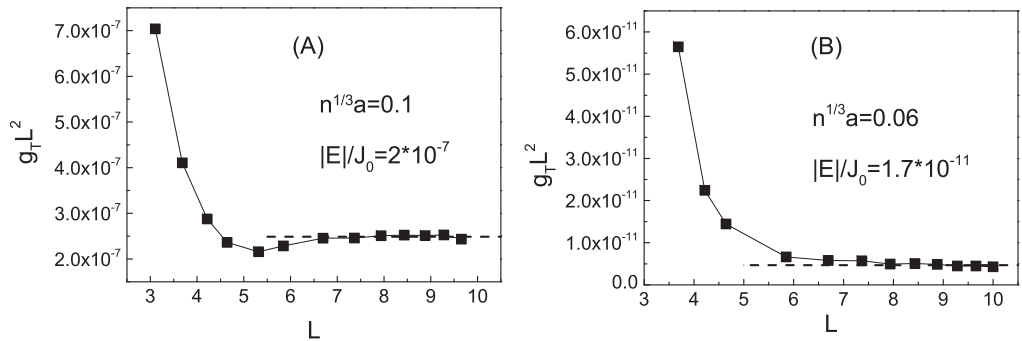}
        \caption{The dependence $g_T(L)$ for the Hamiltonian (\ref{Hmatrix}) with $n^{1/3}a=0.1$ (A) and $n^{1/3}a=0.06$ (B) for the energies near the maximum of the spectrum. The dependence is compared with hypothesis of the diffusive motion. }
    \label{fig:ferdeloc}
\end{figure}

To understand if the localization threshold for the spin problem is smaller than for the Lishitz problem or if there is no localization in the spin problem, we show the dependence $g_T$ on $L$ compared with delocalization hypothesis for smaller values of $n^{1/3}a$: $n^{1/3}a=0.1$ and $n^{1/3}a=0.06$. The corresponding results are shown on the figure \ref{fig:ferdeloc}. At least for the considered parameters there is always a part of the spectrum that corresponds to the delocalized excitations. It contains at least $10\%$ of the energy spectrum. It appears that there is no complete localization in the Heisenberg exchange problem.

Let us discuss why the results for spin Hamiltonian (\ref{Hmatrix}) are different from the results for the Lifshitz localization Hamiltonian (\ref{HLif}). At the first glance the problems are quite similar, in both cases, there is a strong exponential disorder in the non-diagonal terms of the Hamiltonian. The Hamiltonian (\ref{Hmatrix}) also contains the diagonal terms that look like the additional energy disorder (so at first glance the problem (\ref{Hmatrix}) should be ``more localized" than the Lifshitz problem). However the diagonal terms of Hamiltonian (\ref{Hmatrix}) are correlated with the non-diagonal terms leading to the opposite result: there is no localization of the whole spectrum in the problem (\ref{Hmatrix}).

It is possible to show that the Hamiltonian (\ref{Hmatrix}) has at least one delocalized state irrespectively to the degree of disorder. Naturally it can be seen that the state
\begin{equation} \label{granted}
\Psi_1 = \frac{1}{\sqrt{N}} \sum_n \psi_n
\end{equation}
is the eigenstate of the Hamiltonian (\ref{Hmatrix}) with energy $E_1 = E_0$. In the expression (\ref{granted})
$N$ is the number of sites and $\psi_n$ is the wavefunction of the spin excitation localized on the site $n$. The delocalized state $\Psi_1$ has a clear physical meaning. The exchange interaction conserves not only the total spin projection on $z$ axis but also the square of the total spin. So among the spin excitations, there is one with $S=N$ and $S_z = N/2-1$. This excitation has energy $E_0$. In the particle representation, it corresponds to the spin excitation delocalized over all the sites.

Up to now we discussed our results in the dimensionless units suitable for numerics. However, the small number of parameters allows us to get the result for physically interesting parameter values. So let us clarify what diffusion coefficients we obtain for realistic parameters of organic materials or doped semiconductors. We also would like to compare our results with the results of \cite{Yu} so we take the parameters from this study. We consider the localization length $a=1\, nm$, dielectric constant $\epsilon=2$ (that define $J_0 =0.6\,$eV) and consider the concentrations from $10^{17}cm^{-3}$ to $10^{19}cm^{-3}$.

\begin{figure}[htbp]
    \centering
        \includegraphics[width=0.5\textwidth]{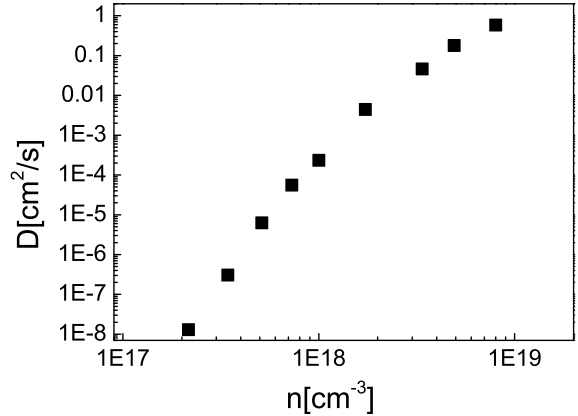}
        \caption{The diffusion coefficient for delocalized states near the maximum of the energy spectrum for different electron concentrations and $a=1\,nm$.  }
    \label{fig:DifAll}
\end{figure}

The diffusion coefficients corresponding to the maximum of the density of states for the discussed parameters are shown on the figure \ref{fig:DifAll} as a function of concentration. The dependence of these coefficients on the energy is always qualitatively similar to the one shown on Fig. \ref{fig:DaLoc}. The diffusion coefficient near the maximum of the density of states is less but comparable with the diffusion coefficient at the maximum. It decreases with the increase of the absolute value of excitation energy $|E|$. At some threshold value of the energy, it tends to zero and the excitations with larger $|E|$ are localized. The relative part of the delocalized excitations decreases with decreasing concentration. It is approximately equal to $25\%$ for $n \sim 10^{19} cm^{-3}$. For small concentrations $n \sim 10^{17} cm^{-3}$ the relative part of the delocalized excitations is $\approx 10\%$. The obtained diffusion coefficients near the maximum of the density of states are by the order of magnitude similar to the diffusion coefficients calculated in \cite{Yu}.

\section{Localization due to the random effective magnetic field and due to the anisotropic exchange interaction.}
\label{sect5}

The absence of the localization in spin systems with Hamiltonian (\ref{Hmatrix}) depends on the exact form of relation between diagonal and non-diagonal matrix elements. In this section we discuss the mechanisms that can change this relation even in the system of spins with exchange interaction.

One of these mechanisms appears when the magnetic field in not perfectly uniform. It also can be related to the hyperfine interaction with nuclear spins that can be described as random on-site effective magnetic fields.
Let us assume that a small non-uniform field ${\bf B}_2({\bf r})$ is added to the uniform magnetic field ${\bf B}$.  The effect of this magnetic field on the exchange interaction elements $J_{nm}$ can be neglected, however ${\bf B}_2$ adds a diagonal term to the Hamiltonian (\ref{Hmatrix}) $\Delta H_{ii} = \mu_B g B_{2z}({\bf r}_i)$. We consider the small correlation length of the random magnetic field, so $\Delta H_{ii}$ are uncorrelated for different sites. Therefore we add a random energy $\Delta E_i = \mu_B g B_{2z}({\bf r}_i)$ to each site, that is selected in the interval $-{\cal E}.. {\cal E}$.

\begin{figure}[htbp]
    \centering
        \includegraphics[width=1.0\textwidth]{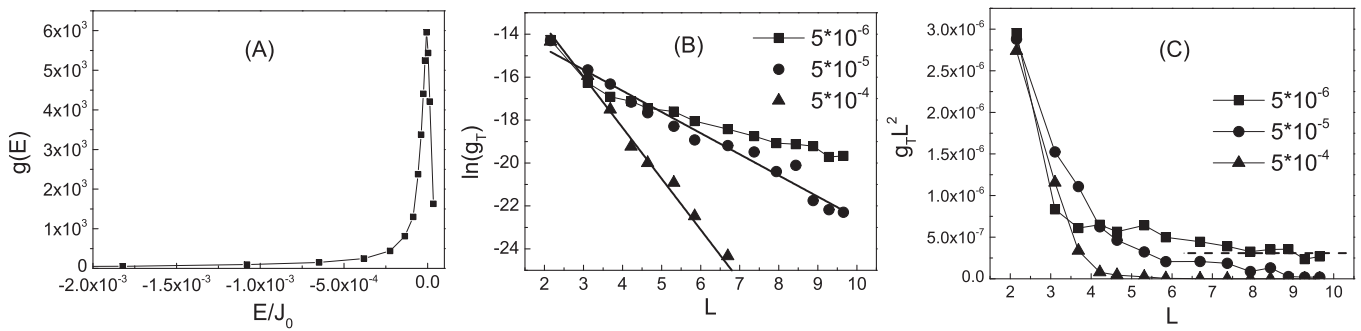}
        \caption{(A) The density of states for the spin Hamiltonian with random energies ${\cal E} = 5\cdot10^{-5} J_0$ and $n^{1/3}a = 0.1$.
        (B,C) The level curvatures $g_T$ for the spin problem with $n^{1/3}a = 0.1$ and different random energies compared
        with the localization and the diffusion hypothesis. Numbers on plots (B) and (C) correspond to the random energy ${\cal E}$ in units $J_0$. }
    \label{fig:REn}
\end{figure}

 Figure \ref{fig:REn} shows the  numerical results for the corresponding spin Hamiltonian with the random effective magnetic field. The density of states is shown on Fig. \ref{fig:REn} (A). Unlike the density of states for the Heisenberg interaction Hamiltonian it contains some excitations with $E>0$. However the maximum of the density of states at $E=0$ is still visible. It corresponds to the maximum diffusion coefficient when the system contains delocalized excitations. The level curvatures for this case are shown on Fig. \ref{fig:REn} (B,C). $g_T$ for energies close to zero (near the maximum of the spectrum) are shown and compared with diffusion and localization hypothesis. For small random energies, the behavior of the level curvature is clearly diffusive. However when the size of the random magnetic field becomes larger than some threshold value the states are localized. The value of ${\cal E}$ that corresponds to the transition ${\cal E} \sim 10^{-5}J_0$ is to the order of magnitude similar to the energy that separates the localized and the delocalized states for the spin Hamiltonian for $n^{1/3}a=0.1$ without random energy.

Spin localization due to the random magnetic field or the hyperfine interaction with nuclear spins is similar to the Anderson localization of electrons. The random magnetic field adds a random on-site energy that is not correlated with exchange integrals. When these random energies exceed the exchange interaction the spin excitations become localized.

The discussed relation between diagonal and non-diagonal Hamiltonian terms is also dependent on the Heisenberg form of the exchange interaction $J_{ij}{\bf S}_i {\bf S}_j$. However in the presence of the spin-orbit interaction it can be altered. The spin-orbit interaction results in the spin rotation around effective magnetic field during the hop. The exchange interaction with respect to this rotation  can be described with Hamiltonian \cite{Kavokin-exc}
\begin{equation}\label{anisH}
H_{ex} = \sum_{ij} J_{ij} \Bigl[ (\widehat{\bf S}_i\widehat{\bf S}_j) \cos\gamma + \frac{(\widehat{\bf S}_i {\bf b})(\widehat{\bf S}_j {\bf b})}{b^2} (1-\cos\gamma) + \frac{{\bf b}}{b} \cdot [\widehat{\bf S_i} \times \widehat{\bf S_j}]\sin\gamma \Bigr].
\end{equation}
Here ${\bf b}$ is the vector of anisotropy and $\gamma$  is the angle of spin rotation during the hop. The first two terms in (\ref{anisH}) correspond to the symmetric anisotropic exchange and the last term is the Dzyaloshinskii-Moriya interaction.

Let us discuss a case when anisotropy vector ${\bf b}$ is always aligned along $x$-axis, while the magnetic field is along $z$ axis. In this case, Dzyaloshinskii-Moriya interaction does not mix the states with the same projection of the total spin and does not contribute to the effective one-particle Hamiltonian. The one-particle Hamiltonian, in this case, appears to be real and can be studied with the twisted boundary conditions method.
\begin{equation}
\label{HmatrixAnis}
H_{ij} = J_{ij}\frac{1+\cos\gamma}{4}, \quad H_{ii} = E_0 - \frac{1}{2}\sum_{j\ne i} J_{ij}\cos\gamma.
\end{equation}
We consider a simple case when the angle $\gamma$ is the same for all hops. Note that contrary to $J_{ij}$ it does not depend exponentially on hopping distance even in real systems.

\begin{figure}[htbp]
    \centering
        \includegraphics[width=1.0\textwidth]{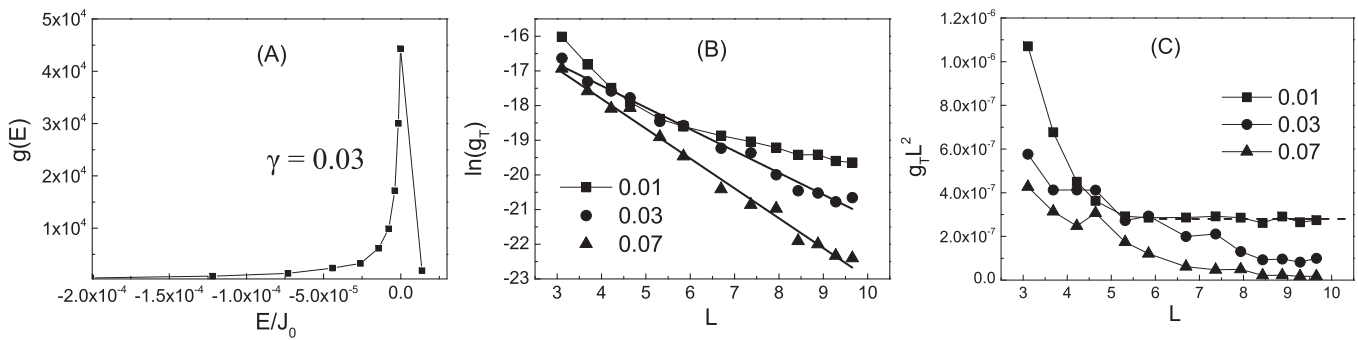}
        \caption{(A) the density of states for the Hamiltonian (\ref{HmatrixAnis}) with $n^{1/3}a = 0.1$ and $\gamma = 0.03$. (B,C) The level curvatures $g_T$ near the maximum of the density of states for the spin problem with $n^{1/3}a = 0.1$ and the anisotropic exchange interaction. Numbers on plots correspond the anisotropy parameter $\gamma$. }
    \label{fig:anis}
\end{figure}

For $\gamma=0$ the Hamiltonian (\ref{HmatrixAnis}) is reduced to the discussed Hamiltonian (\ref{Hmatrix}). For $\cos\gamma=0$ it becomes similar to the Hamiltonian of the Lifshitz problem (\ref{HLif}). So it is natural to assume that spin transport properties for small $na^3$ are dependent on $\gamma$.

The numerical results for the anisotropic exchange Hamiltonian (\ref{HmatrixAnis}) are shown on the figure \ref{fig:anis}. The density of states for $n^{1/3}a = 0.1$ and $\gamma=0.03$ is shown on Fig. \ref{fig:anis} (A). Similarly to the density of states for the spin problem with random magnetic fields there are some spin excitations with positive energies but the maximum at $E=0$  is clearly visible. The level curvatures $g_T$ at the maximum of the density of states are shown on Fig. \ref{fig:anis} (B,C). The localization parameter is $n^{1/3}a=0.1$. For very small anisotropic parameter $\gamma=0.01$  the states near the maximum of the spectrum are delocalized and correspond to the diffusion coefficient $D \sim 2 \cdot 10^{-7}J_0 /\hbar n^{2/3}$. However for larger anisotropy $\gamma \ge 0.03$ all the excitations are localized. We believe that the anisotropic form of the exchange interaction re-establish the physics of the Lifshitz localization and for any finite anisotropy parameter $\gamma$ the localization occurs for sufficiently small $n^{1/3}a$.

The considered situation when the anisotropy axis is the same for all electron pairs can correspond to localized states on the impurities in a doped crystal. The anisotropy axis ${\bf b}$, in this case, is controlled by the symmetry of the crystal matrix. In amorphous organic materials, the spin-orbit interaction originates due to the random orientation of the molecules. So the different pairs of sites should correspond to the different anisotropy axes $\bf b$. In this case, the one-excitation Hamiltonian appears to be complex and can not be directly related to an electron Hamiltonian in some effective electrical potential. It does not allow us to apply the twisted boundary conditions method. However, other numerical methods, such as the calculation of the inverse participation ratio, demonstrate that the anisotropic exchange interaction with random axis $\bf b$ has the same effect as the anisotropic exchange interaction (\ref{HmatrixAnis}). It also leads to the localization of the spin excitations for sufficiently large $\gamma$. The corresponding calculations are presented in appendix 2.

\section{Qualitative description of the localization \label{sect-Qualit}}
\label{sect6}

\begin{figure}[htbp]
    \centering
        \includegraphics[width=0.75\textwidth]{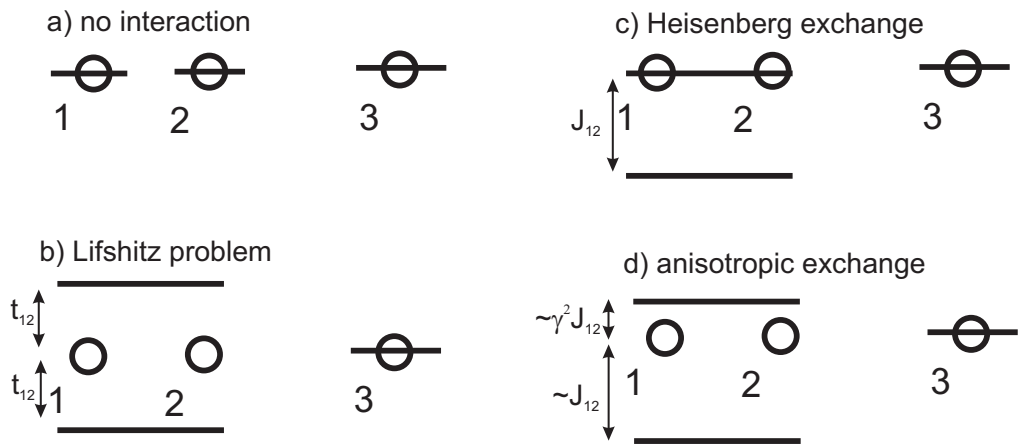}
        \caption{Three sites with different types of interaction. }
    \label{fig:disc}
\end{figure}

Let us discuss qualitatively why the spin Hamiltonian (\ref{Hmatrix}) lacks the Lifshitz localization (although it is quite similar to the Hamiltonian of the Lifshitz problem) and why the anisotropic form of the exchange interaction re-establishes the localization. The minimal model  of the Lifshitz localization involves three sites (see fig \ref{fig:disc}). The sites $1$ and $2$ are separated with the distance $r_{12}$ that is smaller than the distance $r_{23}$. Because the dependence of the exchange integrals on distance is exponentially strong the overlap integral $t_{12}$ is much larger than $t_{23}$.

Without the overlap integrals each site has an electron state with zero energy (fig \ref{fig:disc} a). If one considers the overlap integral $t_{12}$ the energy states $1$ and $2$ form the symmetric combination with energy $-t_{12}$ and the antisymmetric combination with energy $t_{12}$ (fig \ref{fig:disc} b). Therefore the electron can effectively move between sites $1$ and $2$. However the intermixture of the two discussed states with the electron state $3$ is weak $\propto t_{13}/t_{12}$. For small localization parameter $n^{1/3}a \rightarrow 0$ the characteristic size of $t_{13}/t_{12}$ tends to zero leading to the electron localization on a pair of sites. This Lifshitz localization is somewhat similar to the Anderson localization that appears when the system includes the random energy ${\cal E}$ that is larger than the overlap integrals $t_{ij}$. However in the Lifshitz problem the electron is localized on a pair of sites,   the overlap integral $t_{12}$ inside the pair plays the role of the random energy, and the overlap integral $t_{23}$ between the pair and one of the more distant sites plays the role of the overlap integrals in the Anderson problem.

Let us now discuss the Hamiltonian (\ref{Hmatrix}) in terms of the three sites model (fig \ref{fig:disc} c). In this case the states with the flipped spin on sites $1$ and $2$ are split into levels that correspond to the singlet and to the triplet with zero total z-projection. The singlet state has the energy $-J_{12}$ and is effectively separated from the site $3$. However the triplet state has zero energy and can effectively intermix with other sites. Note that the ground state also corresponds to the triplet state of spins on the sites $1$ and $2$ but with $z$-projection with the total spin along the magnetic field. It does not matter how small is the exchange integral $J_{23}$, it can effectively mix the triplet state on sites $1$ and $2$ with the site $3$. It results in the absence of localization in the Heisenberg exchange problem.

However if the exchange interaction has the anisotropic part (fig \ref{fig:disc} d) it does not conserve the total spin and the states appearing due to the exchange between sites $1$ and $2$ cannot be rigorously classified as singlet or triplet. Both states are shifted from zero energy (although for small exchange anisotropy the energy of one of the states is $\propto \gamma^2$). Therefore the localization appears for small $n^{1/3}a$. However the condition for the localization is more strict: $J_{23} \ll \gamma^2 J_{12}$ compared to the analogous condition $t_{23} \ll t_{12}$ in the Lifshitz problem.

Let us note that the discussed explanation can be extrapolated to the situation of the arbitrary magnetization including zero mean magnetization. When the exchange interaction has Heisenberg form the pair of sites 1 and 2 can change z-projection of their total magnetic moment without changing the energy of their exchange interaction. Thus, even small interaction with the site 3 can lead to a spin transition from the pair 1-2 to the site 3 or vice versa. It happens due to the degeneracy of the triplet state of a pair of spins. When the exchange interaction is sufficiently anisotropic the spin states of a pair of sites cannot be classified as triplet or singlet. So the interaction with a site 3 cannot change the spin state of a pair 1-2 when $J_{13}\ll \gamma^2 J_{12}$. In terms of the spin excitations, the situation when the number of flipped spins is not small corresponds to a large concentration of the spin excitations comparable to the number of sites. It leads to the importance of the interaction between spin excitations. We are not able to make a numerical simulation including this interaction, however, we believe that it does not change the fundamental result: when the exchange interaction has a Heisenberg form some part of the spin excitations is always delocalized. However, the anisotropy of the exchange interaction leads to localization for small $n^{1/3}a$.

 When the spin excitations are delocalized they can provide
  a coherent mechanism of the spin transport competing with the transport provided by the phonon-induced hops. The coherent mechanism should dominate over the hopping spin transport at least at low temperatures when the phonon-induced hops are suppressed. Therefore the spin transport can be in principle decoupled for hopping systems.

When the exchange interaction has the anisotropic part or the non-uniform magnetic field or the hyperfine interaction are included into the system all the spin excitations are localized for sufficiently small $n^{1/3}a$. In this case the spin transport is impossible without the interaction with some thermal bath. Phonons cannot provide an effective thermal bath for the exchange excitations due to the small spin-phonon interaction. We believe that a more effective thermal bath can be provided with electron hops. If the localization length of the spin excitations sufficiently exceeds the electron localization length the spin transport can be controlled by a combined process. In this case one electron hop can lead to the exchange-induced re-arrangement of spins in a relatively large area.

\section{Spin-transport due to the exchange interaction and the spin-valve effect}
\label{sect7}

To understand the impact of the discussed spin excitations on the spin transport let us consider a macroscopic
sample with non-interacting spin excitations described with Hamiltonian (\ref{Hmatrix}).  In a macroscopic sample it is
possible to introduce a probability $f_{ex}(E,{\bf r})$ to find a spin excitation with energy $E$ at the position ${\bf r}$. This probability is assumed to be averaged over small but macroscopic volume near the coordinate ${\bf r}$. The finite diffusion coefficient $D(E)$ obtained in the numeric experiment mean that on a macroscopic scale the dependence of $f_{ex}(E,{\bf r})$ on the coordinate ${\bf r}$ leads to appearance of the current of spin excitations with energy $E$,
$j_{ex}(E) = D(E) g_{ex}(E) \nabla f_{ex}(E,{\bf r})$. Here $g_{ex}(E)$ is the density of states of the spin excitations with energy $E$.
Without the external thermal bath the space-integrals $\int f_{ex}(E,{\bf r}) d{\bf r}$ are conserved because no energy relaxation is included into the system. The thermal bath relevant to our problem can be related to the electron hops as discussed in the section \ref{sect-Qualit}. The electron hops can change the exchange interaction coefficients $J_{ij}$ and change the energy distribution of spin excitations but cannot flip spin and conserve the total excitation number. We include this bath phenomenologically introducing the time $\tau_{bath}$ of equilibration of the energy distribution of the spin excitations. It allows us to give a diffusion equation \cite{Dif1,Dif2} for the spin excitations.
\begin{equation}\label{dif-E}
\frac{\partial f_{ex}(E,{\bf r})}{\partial t} = D(E) \nabla^2 f_{ex}(E,{\bf r}) - \frac{1}{\tau_{bath}} (f_{ex}(E,{\bf r}) - f_{ex}^{(eq)}(E,{\bf r}) ).
\end{equation}
Here $f_{ex}^{(eq)}(E,{\bf r}) = n_{ex}({\bf r}) e^{-E/T} / \int e^{-E/T} g_{ex}(E) dE$ is the equilibrium energy distribution of spin excitations. $n_{ex}({\bf r})$ is the concentration of the excitations at the coordinate ${\bf r}$. $f_{ex}^{(eq)}(E,{\bf r})$ corresponds to the Boltzmann distribution because the concentration of spin excitation was considered to be small in the numerical experiment.

In what follows we will be interested in the total polarization of electrons at the coordinate ${\bf r}$. It can be expressed in terms of concentrations of the spin-up electrons $n_{\uparrow}({\bf r})$ and spin-down electrons $n_{\downarrow}({\bf r})$ as $n_{\uparrow}({\bf r})- n_{\downarrow}({\bf r})$.  When there is no spin excitations in the system all spins are polarized and $n_{\uparrow}({\bf r})- n_{\downarrow}({\bf r}) = -n({\bf r})$ where $n({\bf r})$ is the total electron concentration. Each excitation correspond to a flipped spin so a finite number of excitations mean a finite number of up-spins. The relation of the spin excitations to the spin polarization is
\begin{equation} \label{exi-pol}
n_\uparrow({\bf r}) - n_\downarrow({\bf r}) = -n ({\bf r}) + 2 n_{ex} ({\bf r})
\end{equation}

\begin{figure}[htbp]
    \centering
        \includegraphics[width=0.5\textwidth]{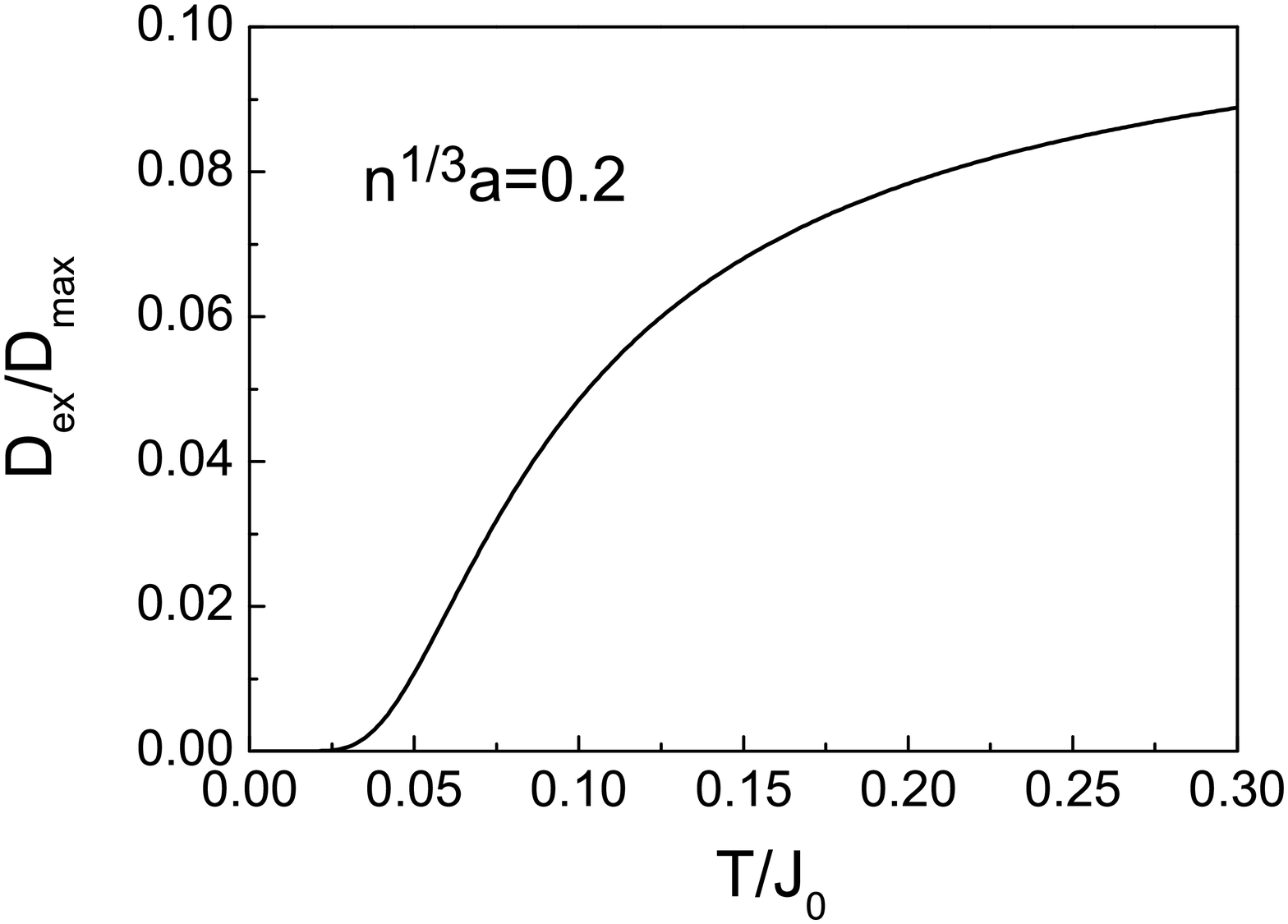}
        \caption{The averaged spin diffusion coefficient for the exchange mechanism of spin diffusion at different temperatures. }
    \label{fig:DexDmax}
\end{figure}

The equation (\ref{exi-pol}) shows that the current of spin excitation $j_{ex} = \int j_{ex}(E) dE$ contribute to the spin current. Expression (\ref{dif-E}) leads to the appearance of the two contributions to the spin excitation current: it can be created by the gradient of  the total number of excitations $\nabla n_{ex}({\bf r})$ or by the gradient of energy distribution of the excitations. When the probabilities $f_{ex}(E,{\bf r})$ are close to their equilibrium values these two contributions can be considered separately. In the present study we restrict ourselves with the situation when the boundary conditions correspond to the equilibrium energy distribution of spin excitations near the edges of the sample. It means that spin injection from the contacts creates a non-equilibrium concentration of spin excitations $n_{ex}({\bf r})$, however the energy distribution of additional excitations is of equilibrium form. In this case we can neglect the non-equilibrium part of energy distribution of the excitations and relate the current $j_{ex}$ only to the gradient of the concentration of spin excitations
\begin{equation}
j_{ex} = D_{ex} \nabla n_{ex}({\bf r}), \quad D_{ex} = \frac{\int D(E) e^{-E/T} g_{ex}(E) dE}{ \int e^{-E/T} g_{ex}(E) dE}.
\end{equation}
Here $D_{ex}$ is the diffusion coefficient of the spin excitations averaged with Botzmann distribution.
The relation of $D_{ex}$ to the maximum spin excitation diffusion coefficient $D_{max}$ corresponding to the excitations with energy $E_0$ is shown on the figure \ref{fig:DexDmax} for $n^{1/3}a = 0.2$.
 Note that even when the temperature is larger than the width of the distribution of spin excitation energies (it is still considered to be small compared to the Zeeman energy in the external magnetic field), $D_{ex}$ is smaller than the diffusion coefficient near the maximum of spin excitation density of states $D_{max}$ shown on figure \ref{fig:DifAll}.   When the temperature tends to zero $D_{ex}$ also tends to zero because the spin excitations with the lowest energy are localized.

Finally we take into account that electron hops not only provide a thermal bath for the spin excitations but also lead to a hopping mechanism of electron and spin transport. We describe this mechanism introducing the hopping diffusion coefficient $D_{hop}$.
\begin{equation} \label{pot2}
j_c =  ge D_{hop} \frac{d\xi_c}{dx}, \quad j_s =  ge (D_{hop} + D_{ex}) \frac{d\xi_s}{dx},
\end{equation}
\begin{equation} \label{j1}
j_c =  const, \quad   \frac{d}{dx} j_s = ge(\xi_s-\xi_s^{(0)})/\tau_{Rel}.
\end{equation}
Here $j_c$ is the charge current, $g$ is the density of electron states at the Fermi level. We introduced the electrochemical potentials for spin-up and spin-down electrons $\xi_\uparrow$ and $\xi_{\downarrow}$. The charge current $j_c$ is related to the averaged electro-chemical potential $\xi_c = (\xi_\uparrow + \xi_{\downarrow})/2$. We also introduced effective spin electrochemical potential $\xi_s = (\xi_\uparrow - \xi_{\downarrow})/2$ that is related to the number of spin excitations as follows $\xi_s = (n_{\uparrow} - n_{\downarrow})/g = (-n + 2n_{ex})/g$. $\xi_s^{(0)}$ is the equilibrium difference of electrochemical potentials for spin-up and spin-down electrons in the applied magnetic field. Equation (\ref{j1}) assumes that all spin-flip processes that lead to the relaxation of the total spin can be described with a single time $\tau_{Rel}$.

The equations (\ref{pot2},\ref{j1}) are derived under assumption that the numerically obtained diffusion coefficients for the spin excitations are valid in the system under consideration, i.e. when $n_{ex}({\bf r}) \ll n({\bf r})$. It can be fulfilled in low-temperature experiments in a magnetic field when temperature is smaller than Zeeman energy.

Although the presented derivation is not valid at room temperature and low magnetic field experiments and is not directly suitable to describe experiments with organic spin-valves, the qualitative arguments (see section \ref{sect-Qualit}) suggest that spin-localization and delocalization picture is not dependent on the number of flipped spins. Therefore the exchange interaction can still contribute to the spin diffusion even when the conditions of the present derivation of equations (\ref{pot2},\ref{j1}) are not fulfilled. Taking into account that the equations (\ref{pot2},\ref{j1}) correspond to the minimal model of introducing different spin and charge diffusion coefficients  we believe that it is interesting to understand how the expressions (\ref{pot2},\ref{j1}) can affect known results on spin-dependent effects if these expressions can be extrapolated to the situations when $n_{ex}({\bf r}) \sim n({\bf r})$, i.e. to high temperatures and small magnetic fields.

Recently it was suggested that the additional spin transport provided by the exchange interaction can be extremely important to the physics of spin-valves \cite{Yu}. It was stated that this spin diffusion removes the conductivity missmatch and make the device insensitive to the perpendicular magnetic field (suppressing the Hanle effect) provided that the spin rotation in the magnetic field is slow compared to the spin diffusion over the device. However, the calculations in \cite{Yu} are focused on the spin injection. The calculations of the magnetoresistance were not provided assuming that the spin-valve magnetoresistance mechanism in organic can be different from the conventional magnetoresistance \cite{Shmidt-SV}. In the present section we show how the fast spin diffusion affects the conventional spin-valve magnetoresistance.

We consider a spin-valve where the spin transport in the normal layer is decoupled from the charge transport. The transport is
described by the equations (\ref{pot2},\ref{j1}) where we consider $\xi_s^{(0)} = 0$ assuming that the equilibrium difference between spin-up and spin-down electro-chemical potentials is small compared to the part of $\xi_s$ related to the spin injection. These equations  should be supplied with boundary conditions that are related to the contacts. In the present study we discuss two situations: when the contact resistances are related to the conductivity of some ferromagnetic materials and when they are related to spin-dependent electron tunneling through the contact barrier. In the later case we assume that the conductivity of the contacts (except the barrier part) is infinite.  We consider the Ohmic contacts because most organic spin-valves operate at low voltages when the whole device is in the linear regime.

It appears that both types of the boundary conditions can be reduced to effective resistance of the contacts $R_{C_1}$ and $R_{C_2}$ corresponding to the left and the right contact and contact spin polarizations $p_1$ and $p_2$. The details related to the boundary conditions and the solution of the transport equations are given in  appendix 3. In both cases the effect of spin polarization in the contacts can be described as the appearance of additional resistance. It includes the part $R_{sv}$ that changes its sign when the magnetization of one of the ferromagnetic contact is flipped.
\begin{equation}\label{Rsv}
R_{sv} = \frac{-4 p_1 p_2 R_{N,s}^{l_{sN}} R_{C_1}R_{C_2}  }{e^{L/l_{sN}}(R_{N,s}^{l_{sN}}  + R_{C_2})(R_{N,s}^{l_{sN}}  + R_{C_1}) - e^{-L/l_{sN}}(R_{N,s}^{l_{sN}}  - R_{C_2})(R_{N,s}^{l_{sN}}  - R_{C_1})}.
\end{equation}
 Here $l_{sN}$ is the spin diffusion length in the normal layer $l_{sN} = \sqrt{(D_{ex} + D_{hop}) \tau_{Rel}^N}$. $\tau_{Rel}^N$ is the spin relaxation time in the normal layer. $L$ is the normal layer thickness. $R_{N,s}^{l_{sN}}$ is the effective spin resistance of the layer of the normal material with the width $l_{sN}$. It is related to the real or "charge'' resistance of the normal layer $R_N$ as $R_{N,s}^{l_{sN}} = R_N  D_{hop} l_{sN}/ (D_{hop} + D_{ex}) L$.

The additional resistance $R_{sv}$ is controlled only by the contact resistances and by the total spin diffusion coefficient $D_{ex} + D_{hop}$ in the normal layer. If the exchange mediated diffusion is large $D_{ex} \gg D_{hop}$, the resistance $R_{sv}$ becomes insensitive to $D_{hop}$. However the resistance measured in the spin-valve experiment is actually the sum $R_{C_1} + R_{C_2} + R_{N} + R_{sv} + R_{pol}$ and includes resistance $R_N$ dependent on the charge diffusion in the normal layer. Here $R_{pol}$ is the part of polarization-dependent resistance that is not changed when one of the contacts changes its magnetization, its expression is different for different boundary conditions (see appendix 3 for details).

\begin{figure}[htbp]
    \centering
        \includegraphics[width=1.0\textwidth]{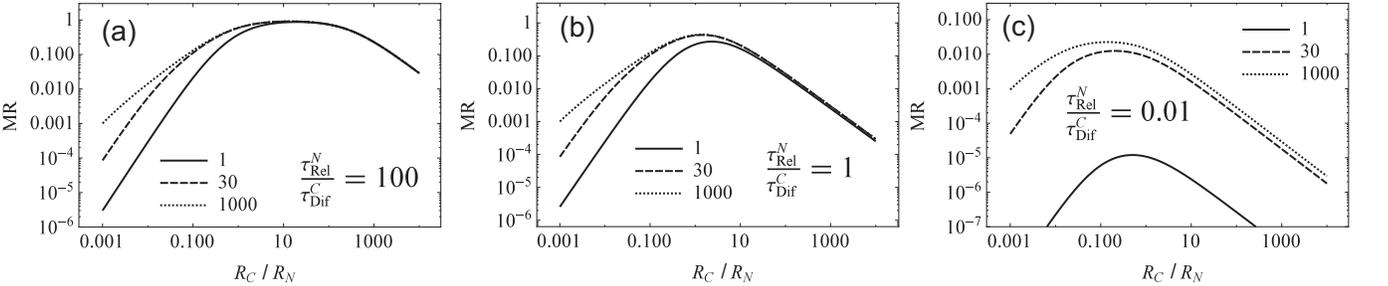}
        \caption{ The dependence of the spin-valve magnetoresistance on the $R_C/R_N$ ratio for different relaxation times $\tau_{Rel}^N$ and different spin diffusion coefficients. Different curves correspond to different ratio $(D_{ex} + D_{hop})/D_{hop}$.}
    \label{fig:kin1}
\end{figure}

Figure \ref{fig:kin1} shows the impact of the difference between charge and spin diffusion coefficients on the spin-valve magnetoresistance defined as $MR = (R_{AP}-R_P)/(R_{AP}+ R_P)$ where $R_{AP}$ is the total device resistivity in the antiparallel configuration and $R_P$ is the resistivity in the parallel configuration. The data is provided for different relations between the contact resistance $R_{C}$ (both contacts are considered to have the same resistivity $R_{C_1}=R_{C_2}$) and the resistance of the normal layer $R_{N}$. Three relaxation times $\tau_{Rel}^N$ in the normal layer are considered on plots (a), (b) and (c). The relaxation times are normalized with the charge diffusion time $\tau_{Dif}^C = 1/D_{hop} L^2$. For each relaxation time we considered three values of $(D_{ex}+D_{hop})/D_{hop}$. The magnitude of the spin-valve magnetoresistance is normalized to the value $MR_{max}=|p_1 p_2|/(2-(p_1^2+p_2^2)/2)$ that correspond to the magnetoresistance in the theoretical limit of the infinite conductivity of the normal layer and the absence of the spin relaxation in the normal layer.

For small contact resistances the value of magnetoresistance is small due to the conductivity mismatch. Without fast spin diffusion the spin-valve resistance in this case is $R_{sv}\sim R_{C}^2/R_{N}$ and the effect is proportional to $R_{C}^2/R_{N}^2$ \cite{Shmidt-SV}. However for high values of $D_{ex}/D_{hop}$ when the effective spin resistance of the normal layer is comparable with the contact resistance $R_{sv} \sim R_C$ and one  order of the smallness is raised. The effect in this case is proportional to the first order of $R_C/R_{N}$. So the fast spin diffusion significantly increase spin-valve magnetoresistance in this case. When $\tau_{Rel}^{N} \ge \tau_{Dif}^C$ (Fig. \ref{fig:kin1} a,b) the magnetoresistance is close to $MR_{max}$ when $R_C$ is comparable with $R_N$. However for very large contact resistance the spin-valve effect is small due to the spin accumulation in agreement with \cite{Fert}. In this case the fast spin diffusion has little effect on the spin-valve because the magnetoresistance is restricted by slow tunneling through the barrier and not by the internal transport in the normal layer. When the spin relaxation in the normal layer is fast $\tau_{Dif}^{N} \ll \tau_{Dif}^C$ (Fig. \ref{fig:kin1} c) the fast spin diffusion always increase the spin-valve magnetoresistance, but it never reaches $MR_{max}$.

\begin{figure}[htbp]
    \centering
        \includegraphics[width=0.75\textwidth]{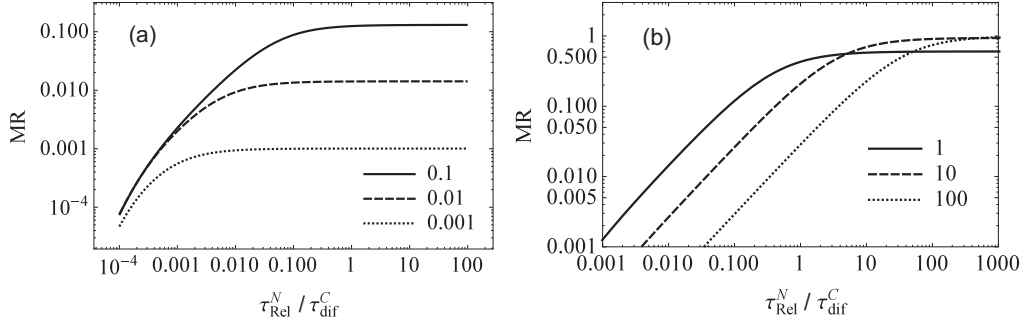}
        \caption{ The dependence of the spin-valve magnetoresistance on the relaxation time in the normal layer $\tau_{Rel}^N$. The numbers on the plots correspond to the ratio $R_C/R_N$. The effective spin diffusion was considered $(D_{ex} + D_{hop})/D_{hop} = 1000$. }
    \label{fig:time}
\end{figure}

It was supposed \cite{Yu} that the fast spin diffusion suppresses the Hanle effect in organic spin-valves when the time of spin diffusion across the sample is smaller than the time of spin rotation in a perpendicular magnetic field. We are not going to give a detailed description of the Hanle effect, however, we assume that the absence of Hanle effect in the perpendicular magnetic field $H_{\perp}$ corresponds to the insensitivity of the spin-valve magnetoresistance to the spin relaxation with times $\tau_{Rel}^N \ge \hbar/ \mu_b H_{\perp}$. To understand if the equations (\ref{pot2},\ref{j1}) lead to such insensitivity we show the dependence of the spin-valve magnetoresistance on the relaxation time in the normal layer $\tau_{Rel}^N$ on figure \ref{fig:time}. It can be seen that that the discussed insensitivity appears when the contact resistivity is small compared to the resistivity of the normal layer. Note that it is the situation when the increase of the spin-valve magnetoresistance due to the fast spin diffusion is the strongest. However in this situation the spin-valve magnetoresistance is always significantly smaller than its maximum value. As it follows from the figure \ref{fig:time} (a) the magnetoresistance is insensitive to spin relaxation times $\tau_{Rel}^{N} \sim 0.01 \tau_{dif}^{C}$ only  when its value is smaller than $1\%$ of $MR_{max}$.

It is interesting to compare the provided results with organic spin-valve experiments exhibiting the absence of the Hanle effect \cite{no-Hanle-1,schmidt_hanle}. The spin-valve magnetoresistance measured in the experiments was of the order of $5\%$. It means that the ratio of magnetoresistance to its maximum value $MR/MR_{max}$ cannot be very small. Taking into account that the maximum magnetoresistance is always less than unity $MR_{max}<1$, it is reasonable to assume $MR/MR_{max} \sim 0.1$.  According to our results shown on the figure \ref{fig:time} it means that the relation of the time of charge diffusion over the sample $\tau_{dif}^C$ to the time of spin relaxation $\tau_{Rel}^N$ should be  $\tau_{dif}^C < 10 \tau_{Rel}^N$.  However, the effective spin relaxation time corresponding to the time of spin precession in the applied perpendicular magnetic field was rather small $\tau_{Rel}^N \sim 10\,$ns. Therefore the time of the charge diffusion over the sample should also be small $\tau_{dif}^C \lesssim 100\,$ns. The thickness of the organic layer in the discussed experiments was $L \sim 100\,$nm. Therefore the charge diffusion coefficient corresponding to $\tau_{dif}^C \lesssim 100\,$ns can be estimated as $10^{-3}cm^2/s$. It is much larger than the charge diffusion coefficient in organics estimated from experiments \cite{Alq3-dif} or the one considered in \cite{Yu}. It means that although our results support the idea that spin and charge transport can be decoupled in organics, this decoupling alone cannot describe all the puzzling experimental results in organic spin-valves.

\section{Conclusion}
\label{sect8}

We studied the problem of excitations in the system of localized spins with exchange interaction and exponentially broad distribution of the exchange integrals in a strong applied magnetic field. The problem was reduced to an effective one-electron problem and studied with the  twisted boundary conditions method.

Although the effective one-electron problem is quite similar to the Lifshitz localization problem, it lacks the complete localization. A part of the spectrum always corresponds to delocalized states that can be described with a diffusion coefficient. The diffusion coefficient exponentially decreases with the decrease of the parameter $n^{1/3}a$.
The appearance of random magnetic or hyperfine on-site fields leads to the Anderson localization of the spin excitations. Anisotropic part of the exchange interaction leads to the Lifshitz localization of the excitations.

The delocalized spin excitations can provide an additional mechanism for the spin transport and make the spin diffusion coefficient larger than the charge diffusion coefficient. This result support the idea that the spin transport in organic spin-valves can be decoupled from the charge transport.
However this decoupling cannot by itself describe all the puzzling experiments with organic spin-valves.

The author is grateful to V.V. Kabanov, V.I. Kozub and M.M. Glazov for many fruitful discussions. The study was partially supported by RFBR project No.~16-02-00064.

\section{Appendix 1: Lifshitz localization problem}

In the main text we compare the exchange interaction Hamiltonian (\ref{Hmatrix}) with the Hamiltonian of the Lifshitz localization (\ref{HLif}). In this appendix we present the results of our numeric computations for the Lifshitz model. This model can be considered as a test for our methods. Also we want to note that although the Lifshitz localization is known since 1965 and was used to discuss several physical systems \cite{Lif-Exp,Lif-Exp2}, we are not aware of any detailed numerical studies of this problem. Thus we believe that the results on this problem can be interesting by themselves.

\begin{figure}[htbp]
    \centering
        \includegraphics[width=1.0\textwidth]{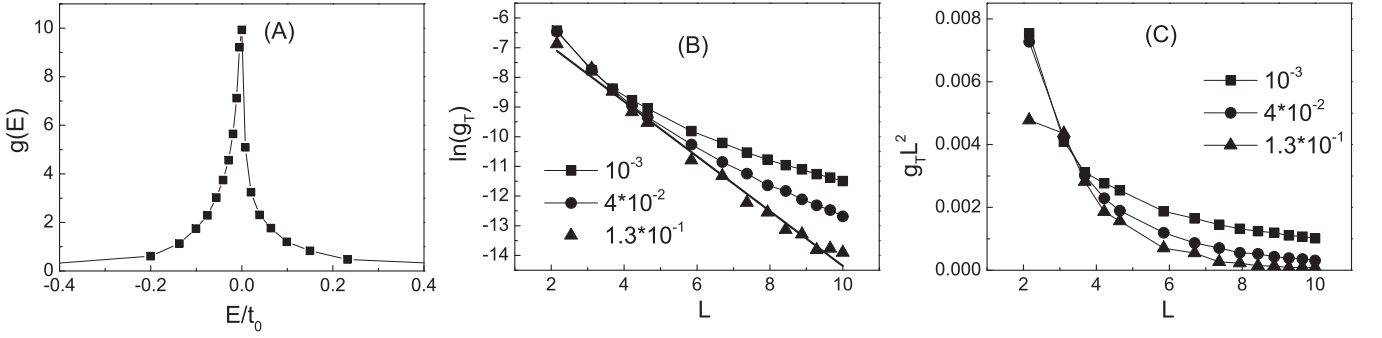}
        \caption{The energy spectrum (a) and  $g_T$ for the Hamiltonian (\ref{HLif}) with $n^{1/3}a= 0.2$.  $g_T$ is compared with two assumptions: of localization (b) and of diffusive motion (c). The numbers on plots (b) and (c) correspond to the energy in units $t_0$. }
    \label{fig:Lif02}
\end{figure}

The Hamiltonian of the problem (\ref{HLif}) is discussed in the main text.
The density of states corresponding to this Hamiltonian for $n^{1/3}a = 0.2$ is shown on fig. \ref{fig:Lif02}(a). It has a sharp maximum near $E=0$ and long power law tails that also contain a large part of the energy levels.

On figure \ref{fig:Lif02} (b) and (c) we show the dependence of the level curvature  $g_T$  on the size of the numerical sample $L$ measured in units $n^{-1/3}$.  The three energies are considered on fig. \ref{fig:Lif02}. The energy $E = 10^{-3}t_0$ lies near the maximum of the spectrum, $E =1.3\cdot 10^{-1}t_0$ is in the tail, while the energy $E = 4\cdot 10^{-2}$ lies somewhere in between.
One can see that the level curvature for $E= 10^{-3}t_0$ clearly corresponds to delocalization while the curvature $g_T$ for $E =1.3 \cdot 10^{-1}t_0$ has the exponential dependence on $L$ indicating the localization with localization distance $\l_{loc} \approx 1.1 n^{-1/3}$.

\begin{figure}[htbp]
    \centering
        \includegraphics[width=1.0\textwidth]{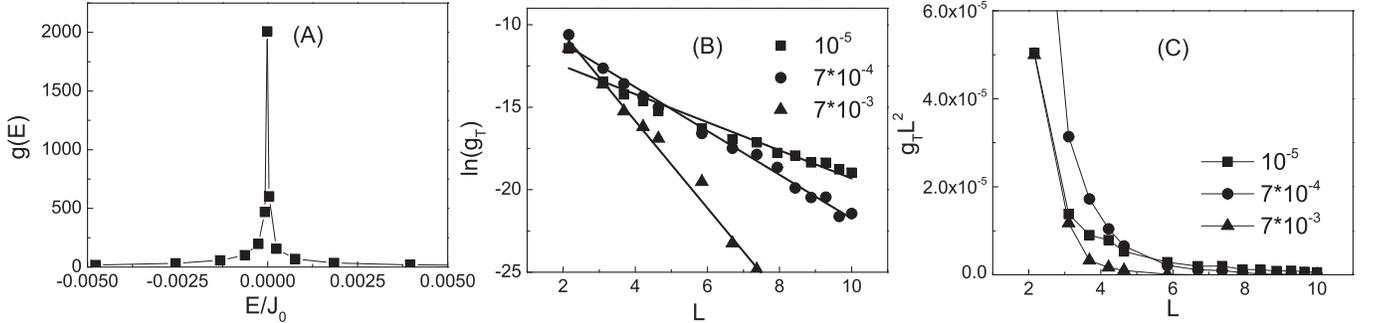}
        \caption{The energy spectrum (a) and  $g_T$ for the Hamiltonian (\ref{HLif}) for $n^{1/3}a= 0.1$.  $g_T$ is compared with two assumptions: of localization (b) and of diffusive motion (c). The numbers on plots (b) and (c) correspond to the energy in units $t_0$. }
    \label{fig:Lif01}
\end{figure}

 Similar numerical results for the smaller localization length $a = 0.1 n^{-1/3}$ are shown on figure \ref{fig:Lif01}. The energy spectrum has similar form as in the case $n^{1/3}a = 0.2$ with sharp maximum and power law tails. However for all the considered energies the $L$ dependence of $g_T$ corresponds to the localization, although the localization distance is larger near the maximum of the density of states than in the tails of the spectrum.

To obtain  the threshold value of $n^{1/3}a$ that correspond to the Lifshitz localization one should slowly change $a$ and find the value $a_{th}$ that correspond to appearance of delocalized states. Following this procedure we obtained $n^{1/3}a_{th} = 0.13 \pm 0.02$. Let us note that in conventional materials (for ex. in doped semiconductors) the metal-insulator transition occurs at larger values $n^{1/3}a \ge 0.2$ \cite{Mott-Davis}. It is usually Anderson or Mott transition. The provided calculation show that Lifshitz mechanism is not important for these materials. It should be important only in the materials with very small random on-site energies.

\section{Appendix 2: IPR method for anisotropic exchange interaction with random anisotropy axis}

The Hamiltonian (\ref{anisH}) includes the anisotropic exchange interaction between sites $i$ and $j$ with anisotropy axis ${\bf b}$. In amorphous organic materials the axis ${\bf b}$ is considered to be random. In this case each pair of sites $i$ and $j$  corresponds to an anisotropy axis ${\bf b}_{ij}$. The reduction of (\ref{anisH}) to the one particle Hamiltonian in this case is
\begin{equation}\label{Hmatr-ran-ax}
H_{ij} = J_{ij} \frac{2\cos\gamma + (1-\cos\gamma)(b_{ij,x}^2 + b_{ij,y}^2)/b^2}{4} - \frac{i}{2} J_{ij} \frac{b_{ij,z}}{b}\sin\gamma,
\end{equation}
$$
H_{ii} = E_0 - \sum_{j\ne i} J_{ij} \frac{\cos\gamma + (b_{ij,z}^2/b^2) (1-\cos\gamma)}{2}.
$$
Here $b_{ij,x}$, $b_{ij,y}$ and $b_{ij,z}$ are the projection of the vector ${\bf b}_{ij}$ on the cartesian axes. The vector ${\bf b}_{ji}$ for the backward hop is opposite to the vector ${\bf b}_{ij}$ of the forward hop ${\bf b}_{ij} = -{\bf b}_{ji}$. Note that the Hamiltonian (\ref{Hmatr-ran-ax}) is complex.

To understand if the Hamiltonian (\ref{Hmatr-ran-ax}) leads to the Lifshitz localization similarly to the Hamiltonian  (\ref{HmatrixAnis}) we calculate the inverse participation ratio for excitations near the maximum of the density of states for different anisotropy types.
The inverse participation ratio (IPR) of a wavefunction $\Psi=\sum_n C_n \psi_n$  is defined as
\begin{equation}\label{IPR}
IPR(\Psi) = \sum_n |C_n|^4.
\end{equation}
Here $\psi_n$ is the wavefunction of the excitation localized on the site $n$. IPR is widely used as a tool to study Anderson localization \cite{anderson-review}. For localized states IPR does not depend on the size of the numerical sample $L$ while for delocalized states IPR tends to zero for large $L$.

\begin{figure}[htbp]
    \centering
        \includegraphics[width=0.75\textwidth]{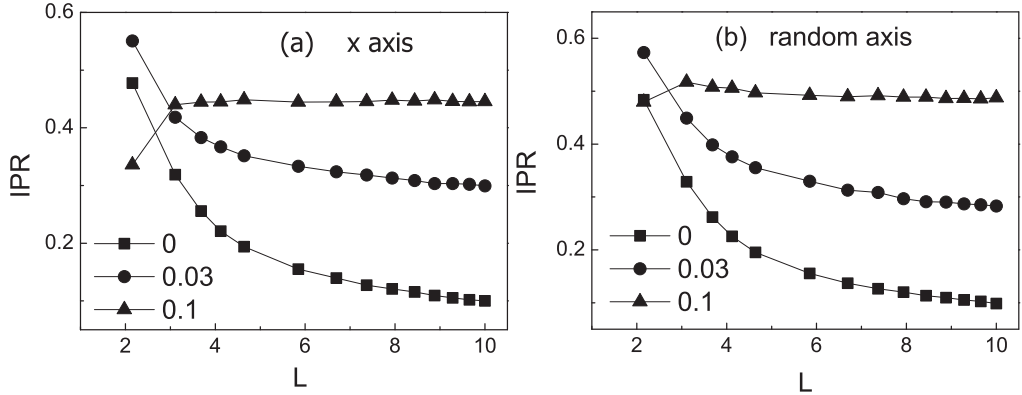}
        \caption{ The inverse participation ratio for anisotropic exchange interaction Hamiltonian with the vector ${\bf b}$ along the x-axis (a) and the random vector ${\bf b}$ (b).  Different curves correspond to different under-barrier spin rotation angles $\gamma$.}
    \label{fig:IPR}
\end{figure}

Figure \ref{fig:IPR} shows the IPR averaged over the excitations near the maximum of the density of states for different underbarrier spin rotation angles $\gamma$. Figure \ref{fig:IPR} (a) corresponds to the  x-axis anisotropy Hamiltonian (\ref{HmatrixAnis}) while figure \ref{fig:IPR} (b) correspond to the Hamiltonian (\ref{Hmatr-ran-ax}) with random axes ${\bf b}_{ij}$. It can be seen that in both cases the anisotropy has a similar effect on IPR. Without the anisotropy of the exchange interaction ($\gamma=0$) IPR tends to zero for large $L$. However for sufficiently large $\gamma$ IPR does not depend on the numerical sample size indicating the localization of the spin excitations.

\section{Appendix 3: the solution of the drift-diffusion equations}

The general solution of the drift-diffusion equations in the normal layer (\ref{pot2},\ref{j1}) is
\begin{equation}\label{soluXi}
\begin{array}{l}
\xi_c(x) = \xi_c(0) + \frac{e j_c}{\sigma_{c}} x, \\
\xi_{s}(x) = C_{1} \exp\left(\frac{-x}{l_{sN}}\right) + C_{2}\exp\left(\frac{x-L}{l_{sN}}\right).
\end{array}
\end{equation}
The coefficients $\xi_c(0)$, $C_{1}$ and $C_{2}$ are determined by the boundary conditions.

In the present study, we consider two types of the boundary conditions. The first one corresponds to the electron dynamics
inside the contact made from a ferromagnetic material.
This dynamics can be described with equations
\begin{equation}\label{FM1}
j_\uparrow = \frac{\sigma^{\uparrow}_{1,2}}{e} \xi_\uparrow',\quad j_\downarrow = \frac{ \sigma^{\downarrow}_{1,2}}{e}\xi_{\downarrow}', \quad j_c'=0, \quad j_s' = \frac{ge}{\tau_{Rel}^{(1,2)}}\xi_s.
\end{equation}
Here $j_\uparrow$ and $j_\downarrow$ stand for the parts of the current carried by the spin-up and spin-down electrons correspondingly.
$\sigma^{\uparrow}_{1,2}$ is the conductivity of the ferromagnetic material for spin up electrons, the bottom index ${1}$ and $2$ correspond to the left and the right contact. $\sigma^{\downarrow}_{1,2}$ is the same for spin down electrons. $\tau_{Rel}^{(1)}$ and $\tau_{Rel}^{(2)}$ are the spin relaxation times in the left and the right contact correspondingly. The boundary conditions related to the ferromagnetic contact are the continuity of spin and charge currents and electro-chemical potentials $\xi_s$ and $\xi_c$ on a ferromagnetic-normal layer interface.

The application of the boundary conditions (\ref{FM1}) leads to the system of equations for  $C_1$ and $C_2$
\begin{equation}\label{a2-c-sys}
\begin{array}{l}
C_{1}\left( \frac{\widetilde{\sigma}_{1}}{el_{1}} + \frac{\sigma_s}{el_{sN}} \right) +
C_{2}e^{-L/l_{sN}}\left( \frac{\widetilde{\sigma}_{1}}{el_{1}} - \frac{\sigma_s}{el_{sN}} \right) = - p_1 j_c,
\\
C_{1}e^{-L/l_{sN}}\left( \frac{\widetilde{\sigma}_{2}}{el_{2}} - \frac{\sigma_s}{el_{sN}} \right) +
C_{2}\left( \frac{\widetilde{\sigma}_{2}}{el_{2}} + \frac{\sigma_s}{el_{sN}} \right) = p_2 j_c.
\end{array}
\end{equation}
Here $\widetilde{\sigma}_{1,2} = 4\sigma_{1,2}^\uparrow \sigma_{1,2}^\downarrow/(\sigma_{1,2}^\uparrow + \sigma_{1,2}^\downarrow)$. The effective thicknesses $l_{1}$ and $l_2$ of the ferromagnetic layers are $l_{1,2} = \sqrt{\tau_{1,2}\widetilde{\sigma}_{1,2}/g_{\mu 1,2}e^2}$, where $g_{\mu 1,2}$ is the density of states in the ferromagnetic contacts. The contact polarizations $p_{1,2}$ are defined as $p_{1,2} = (\sigma_{1,2}^\uparrow - \sigma_{1,2}^\downarrow)/(\sigma_{1,2}^\uparrow + \sigma_{1,2}^\downarrow)$.
The solution of the equation (\ref{a2-c-sys}) together with the general solution (\ref{soluXi}) determines the dependence $\xi_s(x)$ inside the normal layer.

To calculate the device resistivity the difference between electrochemical potential $\xi_c$ on the opposite sides of the device should be found. The change of $\xi_c$ inside the normal layer is always equal to $eL j_c/\sigma_c$. In this sense the resistivity of the normal layer is always equal to $R_N$. Inside the ferromagnetic contacts $\xi_c$ can be described with the expressions
\begin{equation}\label{FM2}
\begin{array}{l}
\xi_c(x) = \xi_c(0) + \frac{ej_cx}{\sigma_1^\uparrow + \sigma_1^\downarrow} - \xi_s(0) p_1 \left( e^{x/l_{1}} -1\right),
\\
\xi_c(x) = \xi_c(L) + \frac{ej_cx}{\sigma_2^\uparrow + \sigma_2^\downarrow} - \xi_s(L) p_2 \left( e^{ -(x-L)/l_{2}} -1 \right).
\end{array}
\end{equation}
Here the first expression corresponds to the left ferromagnetic contact and the second expression corresponds to the right one.

The part of (\ref{FM2}) related to $\xi_s$ leads to the additional contact resistance equal to $(p_2\xi_s(L) - p_1\xi_s(0))/ej_c$. The contact resistance appears to be quadratic in terms of contact polarizations $p_1$ and $p_2$. It can be divided into the two terms: $R_{sv} \propto p_1 p_2$ that changes its sign when one of the contacts flips its magnetization and $R_{pol}$ that depends on the degree of contact polarizations but is insensitive to its sign.
\begin{equation}\label{app-Rsv}
R_{sv} = \frac{-4 p_1 p_2 R_{N,s}^{l_{N}} R_{C_1}R_{C_2}  }{e^{L/l_{N}}(R_{N,s}^{l_{N}}  + R_{C_2})(R_{N,s}^{l_{N}}  + R_{C_1}) - e^{-L/l_{N}}(R_{N,s}^{l_{N}}  - R_{C_2})(R_{N,s}^{l_{N}}  - R_{C_1})}
\end{equation}
\begin{equation}\label{app-Rpol}
R_{pol} = 2\frac{p_1^2 R_{N,s}^{l_{N}} R_{C_1} (R_{N,s}^{l_{N}} \sinh \frac{L}{l_N} + R_{C_2}\cosh \frac{L}{l_N}) + p_2^2 R_{N,s}^{l_{N}} R_{C_2} (R_{N,s}^{l_{N}} \sinh\frac{L}{l_N} + R_{C_1}\cosh\frac{L}{l_N}) }{e^{L/l_{N}}(R_{N,s}^{l_{N}}  + R_{C_2})(R_{N,s}^{l_{N}}  + R_{C_1}) - e^{-L/l_{N}}(R_{N,s}^{l_{N}}  - R_{C_2})(R_{N,s}^{l_{N}}  - R_{C_1})}
\end{equation}
Here $R_{C_{1,2}} = l_{1,2}/\widetilde{\sigma}_{1,2}$ are the effective contact resistances.

The second type of boundary conditions is related to the introduction of spin-dependent barriers related to the interface between
the ferromagnetic and normal layers in the spin-valve.
This spin filtering barriers are described with equations
\begin{equation}\label{bar1}
\Delta\xi_{\uparrow, \downarrow}(0) = \frac{e j_{\uparrow, \downarrow}}{\Sigma_{\uparrow,\downarrow}^l}, \quad \Delta\xi_{\uparrow, \downarrow}(L) = \frac{e j_{\uparrow, \downarrow}}{\Sigma_{\uparrow,\downarrow}^r}.
\end{equation}
Here $\Sigma_\uparrow^l$ and $\Sigma_\downarrow^l$ are the conductivity of the left barrier for spin up and spin down electrons correspondingly. $\Sigma_\uparrow^r$ and $\Sigma_\downarrow^r$ are the same for the right barrier. $\Delta\xi_\uparrow(0)$ and $\Delta\xi_\uparrow(L)$ are the shifts of electro-chemical potential of spin-up electrons at the left and the right barriers correspondingly. The boundary conditions in this case assume that electron and spin current are continuous on the interface and $\xi_s$ outside the tunnel barriers is equal to zero.

The system with the boundary conditions (\ref{bar1}) can be treated in the same way as the system with the boundary conditions (\ref{FM1}). The expression for the contact resistance in this case is
\begin{equation}\label{Rcon-bar}
R_{sv} + R_{pol} = -p_1\frac{1}{\widetilde{\Sigma}_l} \frac{j_s(0)}{j_c} - p_2\frac{1}{\widetilde{\Sigma}_r}\frac{j_s(L)}{j_c}
\end{equation}
where $\widetilde{\Sigma}_{l,r} = 4\Sigma_{l,r}^\uparrow \Sigma_{l,r}^\downarrow /(\Sigma_{l,r}^\uparrow+ \Sigma_{l,r}^\downarrow)$, $p_1 = (\Sigma_{l}^\uparrow - \Sigma_{l}^\downarrow)/(\Sigma_{l}^\uparrow + \Sigma_{l}^\downarrow)$ and $p_2 = (\Sigma_{r}^\uparrow - \Sigma_{r}^\downarrow)/(\Sigma_{r}^\uparrow + \Sigma_{r}^\downarrow)$. Eq.~(\ref{Rcon-bar}) leads to the expression for $R_{sv}$ that is similar to (\ref{app-Rsv}) where the contact resistances are defined as $R_{C_{1,2}} = 1/\widetilde{\Sigma}_{1,2}$. The expression for $R_{pol}$ is however different from (\ref{app-Rpol})
\begin{equation}\label{app-Rpol2}
R_{pol} = -2\frac{p_1^2  R_{C_1}^2 (R_{N,s}^{l_{N}} \cosh \frac{L}{l_N} + R_{C_2}\sinh \frac{L}{l_N}) + p_2^2  R_{C_2}^2 (R_{N,s}^{l_{N}} \cosh\frac{L}{l_N} + R_{C_1}\sinh\frac{L}{l_N}) }{e^{L/l_{N}}(R_{N,s}^{l_{N}}  + R_{C_2})(R_{N,s}^{l_{N}}  + R_{C_1}) - e^{-L/l_{N}}(R_{N,s}^{l_{N}}  - R_{C_2})(R_{N,s}^{l_{N}}  - R_{C_1})}
\end{equation}
The resistance $R_{pol}$ described with expression (\ref{app-Rpol2}) is negative. Note that $R_{pol}$ does not represent any physical resistance, it is just the part of the contact resistance that correspond to the spin polarization. The total contact resistance always stays positive.

\end{document}